\documentclass{jltp}
\usepackage{graphicx}
\usepackage{bm}
\title{Dissipation in Mesoscopic Superconductors with Ac Magnetic Fields}

\author{A. D. Hern\'{a}ndez$^{*,\dag}$, O. Ar\'{e}s$^*$, C. Hart$^*$, 
D. Dom\'{\i}nguez$^\dag$,
\\ H. Pastoriza$^\dag$  and A. Butera$^\dag$}

\address{$^*$ Laboratorio de Superconductividad, Universidad de La
Habana, Cuba
\\$^\dag$ Centro At\'{o}mico Bariloche, 8400
S. C. de Bariloche, Argentina}

\runninghead{A. D. Hern\'{a}ndez {\it et al.}}
{Mesoscopic superconductors}

\begin{document}

\maketitle

\begin{abstract}
The response of mesoscopic superconductors to an ac magnetic field is
investigated both experimentally and with numerical simulations.
We study small square samples with dimensions of
the order of the penetration depth.
We obtain the ac susceptibitity $\chi=\chi'+i\chi''$ at microwave frequencies as
a function of the dc magnetic field $H_{dc}$.
We find that the dissipation, given by $\chi''$, 
has a non monotonous behavior in mesoscopic samples.
In the numerical simulations we obtain that the dissipation increases before the
penetration of vortices and then it decreases abruptly  after vortices 
have entered the sample. This is verified experimentally, where
we find that $\chi''$ has strong oscillations as a function of $H_{dc}$
in small squares of $Pb$.

\end{abstract}

The response of superconductors to an ac magnetic field has been of
interest in the last years\cite{clem}.
The microwave surface impedance and the ac
magnetic susceptibility $\tilde\chi=\chi' + i\chi''$
have been extensively studied. The imaginary part of the 
susceptibility, $\chi''$, is proportional to the dissipation
in the sample. In macroscopic type II superconductors, 
the ac dissipation increases with
magnetic field, proportional to the vortex density.
Very recently, there has been an interest in the study of mesoscopic
superconductos where the sample dimensions are of the order of
the London penetration depth\cite{peeters}. 
 In this work we will show that in mesoscopic superconductors the 
ac dissipation has a non monotonous and oscillating 
dependence with magnetic field.

\begin{figure}[ht]
\includegraphics[width=0.8\linewidth]{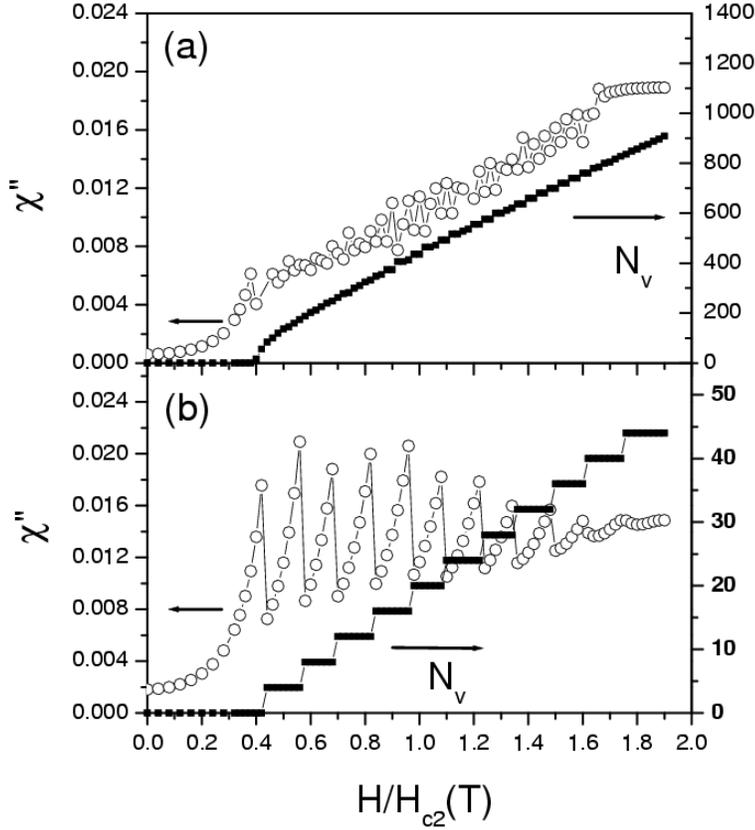}
\caption{ac dissipation, $\chi''$, and number of vortices, $N_v$,
vs. magnetic field $H$. (a) Large sample, $40\lambda\times 40 \lambda$.
(b) Mesoscopic sample, $10\lambda\times 10 \lambda$ }  
\end{figure}
We first show the results of
numerical simulations of mesoscopic squares 
using the  time-dependent Ginzburg-Landau (TDGL)
equations:\cite{kato,ad} 
\begin{eqnarray} 
\frac{\partial \Psi}{\partial t} = \frac{1}{\eta} [(\nabla -
i{\mathbf A})^2 \Psi +(1-T)(1-|\Psi |^2)\Psi ]   \label{tdgl1}\\
\frac{\partial {\mathbf A}}{\partial t} = (1-T)\mbox{Im}[\Psi^* (\nabla -
i{\mathbf A})\Psi] -\kappa^2\nabla \times \nabla \times {\mathbf A}
\label{tdgl2}
\end{eqnarray}
where $\Psi$ and ${\mathbf A}$ are the order parameter and vector potential
respectively and  $T$ is the temperature. 
Lengths are normalized by 
the coherence length $\xi(0)$, times by
$t_0=4\pi \sigma_n\lambda_L^2/c^2$, ${\mathbf A}$ by $H_{c2}(0) \xi
(0)$, $T$ by $T_c$; and  we have 
$\eta=c^2/(4\pi\sigma_n\kappa^2D)$. 
The  boundary
conditions are $(\nabla -i\vec{A})^\perp \Psi =0$ and 
$B_z|_S=(\nabla \times {\mathbf A})_z|_S=H_z^{ext}$. 
We consider an ac+dc magnetic field: 
$H_z^{ext}=H_{dc} + h_{ac}\cos(\omega t)$ with
$h_{ac} \ll H_{dc}$.
The magnetization is
given by $4 \pi M(t)=\langle B(t) \rangle - H(t)$,
with $\langle B(t) \rangle$ averaged over the sample.
The ac magnetic susceptibilities are obtained as $ 
\chi ' = \frac{1}{\pi h_{ac}} \int_0^{2 \pi} M(t)\cos(\omega t)
d(\omega t)$ and
$\chi " = \frac{1}{\pi h_{ac}} \int_0^{2 \pi} M(t)\sin(\omega t)
d(\omega t)$.
We consider 
here $\kappa=2$, $T=0.5$ and the parameter $\eta=12$. 
We use a finite difference discretization scheme\cite{kato,ad}
with $\Delta x= \Delta y=0.5 \xi$ 
and  time step  $\Delta t \le 0.015 t_0$.
Fig.~1(a) shows $\chi" (H_{dc})$ at $\omega= 0.06 \nu_{o}$ obtained in a 
``large'' sample of $40\lambda \times 40\lambda$. We also show
in Fig.1(a) the number of vortices as a function of $H_{dc}$.
At small $H_{dc}$, in the Meissner state, 
$\chi"$  increases continuously with increasing $H_{dc}$.
In this case, the magnetic 
field induces static supercurrents which deplete
the order parameter at the boundary and this 
leads to an increase of dissipation for increasing $H_{dc}$.
Above a penetration field $H_p>H_{c1}$ (there is a
surface barrier\cite{ad}) vortices enter into the sample
and the dissipation increases linearly with the number
of vortices.  
In Fig.~1(b) we show the similar case but for a mesoscopic
square of size $10\lambda \times 10\lambda$.
When the magnetic field is increased above the first
penetration field $H_{p}\equiv H_{p1}$, the 
first vortices enter into the sample, and we see
that there is a discontinuous 
jump in $\chi"$ with a {\it decrease} of dissipation just at $H_{p}$.
Further jumps in $\chi" (H_{dc})$ are present at
the other magnetic fields for vortex penetration, 
$H_{p2},H_{p3},\dots$, where $\chi''$ {\it decreases}. 
The jumps  are followed by a later continuous increase of $\chi" (H_{dc})$
with increasing $H_{dc}$ while the number of vortices $N_v$ remains
fixed. This increase of dissipation at fixed $N_v$ has been
interpreted as due to ``nascent vortices''\cite{ad} which give a 
large fluctuation of $|\Psi|$ at the boundary.
Later, when vortices enter,  
there is a sudden decrease of the ac losses.

\begin{figure}[ht]
\includegraphics[width=0.8\linewidth]{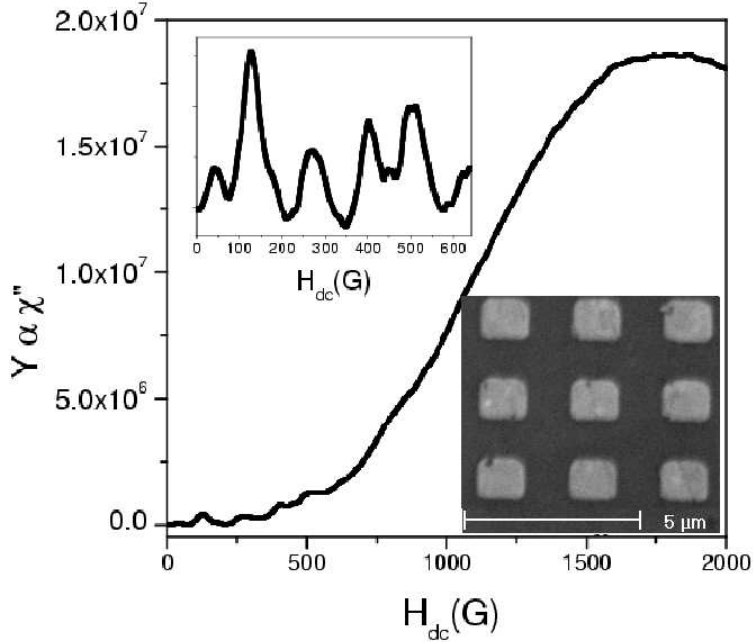}
\caption{Measured ac susceptibility, $Y \propto \chi''$,
 in $1.3\,\mu$ Pb squares
at 9.5 GHz. Lower inset: SEM image of part of the sample.}  
\end{figure}\noindent

We now discuss the experimental measurements.
An array of $200\times250$ Pb squares were deposited by thermal
evaporation on a silicon subtrate. The squares of $L=1.3\,\mu$m on side
and $d=300 \AA\;$ thick where defined by e-bem lithography and a subsequent
lift-off. This thickness $d$ corresponds to an effective 
$\kappa_{eff}=\kappa_{Pb}^2\xi_{Pb}/d\approx 1.3$
and to a length $L\approx 10\lambda_{eff}$.
A SEM image of part of the sample is shown in the lower inset
of Fig.~{2}. 
Magnetic resonance measurements have been made with a 
commercial Bruker ESP 300 spectrometer. Spectra were acquired at
$9.5$ GHz (X-Band) using a TE102 rectangular cavity. 
The external magnetic field was modulated with a 100 kHz ac field of
varying amplitude to obtain the field derivative of the microwave absorption
  ($d\chi''/dH$). 
Low temperature measurements were made
using a continuous flow helium cryostat. 
The sample was positioned inside the cavity with the dc field and the rf field
perpendicular and parallel to the film plane, respectively. 
Spectra features were generally of very weak intensity and
it was necessary to use long acquisition times (several minutes) 
in order to obtain a reasonably good signal to noise
ratio.  In this way, spectra of
$d\chi''/dH$ vs. $H$ were obtained. Fig.~2 shows the integrated
curve, which is proportional to $\chi''$, as function of the
the magnetic field $H$ 
for a temperature $T=4.5K$. We first observe an overall increase of
$\chi''$ with $H$ which is also observed in bulk $Pb$ and it is
due to microscopic mechanisms (not described by the TDGL equations).
An enhancement of the behavior of $\chi''$ at low fields is
shown in the upper inset of Fig.~2. We see that the ac dissipation
has a non monotonous oscillatory behavior as a function of $H$.
This is in qualitative agreement with the theoretical result
of Fig.~1(b). Effects of disorder due to small variations in the
shape of the squares may cause that the oscillations of $\chi''$
are not as uniform as in Fig.~1(b). 
In conclusion, we have observed
that in mesoscopic squares the dependence of ac dissipation with
magnetic field is very sensitive to vortex entrance events.

We acknowledge support from Fundaci\'{o}n Antorchas, SECYT-CITMA
international cooperation agreement, CLAF and ANPCyT.

\end{document}